\documentstyle[12pt,aasms4]{article}  

\begin{document}

\title{A Survey for Outer Satellites of Mars: Limits to Completeness}  
\author{Scott S. Sheppard\altaffilmark{1}, David Jewitt and Jan Kleyna}    
\affil{Institute for Astronomy, University of Hawaii, \\
2680 Woodlawn Drive, Honolulu, HI 96822 \\ sheppard@ifa.hawaii.edu, jewitt@ifa.hawaii.edu, kleyna@ifa.hawaii.edu}

\altaffiltext{1}{Current address: Department of Terrestrial Magnetism,
Carnegie Institution of Washington, 5241 Broad Branch Rd. NW ,
Washington, DC 20015; sheppard@dtm.ciw.edu.}

\begin{abstract}  

We surveyed the Hill sphere of Mars for irregular satellites.  Our
search covered nearly the entire Hill Sphere, but scattered light from
Mars excluded the inner few arcminutes where the satellites Phobos and
Deimos reside.  No new satellites were found to an apparent limiting
red magnitude of 23.5, which corresponds to radii of about 0.09 km
using an albedo of 0.07.

\end{abstract}


\section{Introduction}

The planets have two main types of satellite (Peale 1999).  Regular
satellites have nearly circular, low inclination, prograde orbits.
These satellites probably formed within a disk of gas and dust around
the planet analogous to the disk from which the planets formed around
the Sun.  Irregular satellites have large, eccentric orbits that are
usually highly inclined and even retrograde relative to the equatorial
plane of their planets.  These irregular satellites cannot have formed
by circumplanetary accretion.  Instead, they were probably captured
from heliocentric orbit during the epoch of formation of the planets
(Kuiper 1956; Pollack, Burns \& Tauber 1979).

Permanent satellites are confined to the region of gravitational
influence, known as the Hill Sphere.  The Hill sphere radius is
\begin{equation}
r_H = a_p \left[\frac{m_p}{3M_{\odot}}\right]^{1/3}
\label{eq:hill}
\end{equation}
where $a_p$ and $m_p$ are the semi-major axis and mass of the planet
and $M_{\odot}$ is the mass of the Sun. From Equation~\ref{eq:hill}
and using a mass for Mars of $0.64 \times 10^{24}$ kg, the mass of the
Sun as $1.99 \times 10^{30}$ kg and data from Table~1 we find that
Mars' Hill sphere is about 0.71 degrees in radius which corresponds to
about 1.6 square degrees in area as seen from the Earth during our
observations.

The availability of sensitive, large scale Charge-Coupled Device (CCD)
detectors has refreshed the study of irregular satellites by enabling
a new wave of discovery around Jupiter (Sheppard \& Jewitt 2003),
Saturn (Gladman et al. 2001), Uranus (Gladman et al. 2000) and Neptune
(Holman et al. 2003; Sheppard et al. 2003).  Large fields of view are
needed because the Hill spheres are large.  Sensitivity is needed
because the majority of irregular satellites are small and therefore
faint.  From these new surveys it appears that the four giant planets
possess about the same number of irregular satellites and groupings
with no dependence on the planet's mass when comparing them to the
same limiting size of satellites (Sheppard and Jewitt 2003).  For an
in depth review of irregular satellites see Jewitt, Sheppard \& Porco
(2004).

Mars is known to have two small inner satellites, Phobos and Deimos
(Hall 1878).  The origins of these satellites are still a mystery.
Their orbits are not like the irregular satellites of the giant
planets since they are close to Mars and have nearly zero eccentricity
and inclination.  Mars' satellites, from a dynamical perspective,
appear to have originated not far from their current positions since
it is hard to produce orbits like those of the known Martian
satellites by capture.  On the other hand, the physical properties of
Phobos and Deimos resemble those of C-type asteroids of which most are
found in the outer main belt, not near Mars.  Based on their physical
characteristics they may be captured asteroids.  In addition, the
current orbit of Phobos is short-lived and the satellite will collide
with the planet on a timescale $\sim 10^{8}$ yr (see Burns (1992) and
references therein).

Kuiper (1961) briefly reviewed some early photographic searches for
outer satellites of Mars.  These surveys only reached a photographic
magnitude of about 17 mag (about 1.4 km in diameter assuming an albedo
of 0.05).  The Viking Orbiter 1 spacecraft found no new satellites of
Mars larger than about 0.05 km within the orbit of Phobos (Duxbury \&
Ocampo 1988).  Showalter, Hamilton \& Nicholson (2001) used the Hubble
Space Telescope (HST) to search for possible Mars dust rings created
by Phobos and Deimos that have been predicted by Soter (1971).  They
did not detect any dust rings or unknown satellites down to about 0.1
km in diameter (assuming an albedo of 0.07) in the very inner portions
of Mars space near Phobos and Deimos.  The HST observations have not
been published, but the small field-of-view (a few arcminutes) in any
case allowed them to survey only the very inner portion of the Hill
sphere of Mars.  In this sense, our survey is complementary to theirs.

Only the giant planets have been found to possess abundant irregular
satellites.  Are the distant irregular satellites unique to the giant
planets or does the outermost terrestrial planet, Mars, have them as
well?  Like Jupiter and Neptune, Mars has Trojan asteroids in its
leading and trailing Lagrangian regions which might have been captured
in a similar way as irregular satellites (Marzari et al. 2002).  With
modern wide-field digital detectors on medium class telescopes we can
now survey the space around Mars to much fainter magnitudes than the
early photographic surveys.  In this paper we describe such a survey
in which we used a wide-field digital detector to perform a search of
the Martian Hill sphere for satellites several magnitudes fainter than
previously achieved.

\section{Observations}

We used the Canada-France-Hawaii 3.6 meter diameter telescope atop
Mauna Kea in Hawaii with the MegaCam digital CCD camera for imaging
(Boulade et al. 2003).  The MegaCam has 36 separate CCDs with each
having $2048 \times 4612$ ($13.5 \micron$) pixels, which at prime
focus gives $0.\arcsec 187$ pixel$^{-1}$ scale.  The total number of
pixels is about 340 million and each image was about 710 Megabytes in
size. The field-of-view is $0.96 \times 0.94$ degrees with gaps of
about $13 \arcsec$ between most chips (though a few have gaps of about
$80 \arcsec$).  MegaCam uses a wide field corrector in order to
produce uniform image quality across the field of view.

Images were obtained by staff observers in queue scheduling mode.  An
r' filter based on the Sloan Digital Sky Survey (SDSS) filter set was
used.  The images were bias subtracted and then flat-fielded using
twilight flats.  During exposures the telescope was autoguided
sidereally on bright nearby stars.  The nights were photometric and
the seeing was about $0.\arcsec 7$ during the observations.
Integration times were 100 seconds on July 5 and 120 seconds on July
7.  The observing geometry and characteristics of Mars at the time of
observations are shown in Table~1.

Figure~1 shows the area searched around Mars for possible satellites.
The entire outer Martian Hill sphere was covered on each night of UT
July 5 and 7, 2003.  Four fields were imaged 3 times each on both
nights for a total of 24 images.  On a given night, the images of each
field were separated by about 13.5 minutes giving 27 minutes between
the first and last images.  A computer program was used to identify
objects which were detected in all three images from one night and had
a motion within $\pm 6$ arcseconds per hour of Mars' motion in both
Right Ascension and Declination.  This motion is consistent with
parallactic displacement of an object moving near Mars.  There was no
confusion with main belt asteroids since they tend to stay away from
Mars' orbit and Mars was near perihelion during the observations.

Each night was photometric and Landolt (1992) standards were used to
calibrate the data.  The apparent red limiting magnitude of the survey
was found by placing artificially generated objects with motions
similar to Mars into the survey fields.  The artificial objects had a
range of magnitudes and were matched to the point spread function of
the images in the North-South direction (FWHM $\sim 0.7$ arcseconds)
and were elongated in the East-West direction to simulate the slight
trailing any real satellites would have incurred ($\sim 1.1$
arcseconds) during the exposure.  The $50 \%$ differential detection
efficiency for artificial moving objects in the survey fields was
found at an R-band magnitude of about $23.5$ mags.
Figure~\ref{fig:eff} shows the differential detection efficiency of
the search program versus limiting red magnitude.  The multiple nights
of observations which would only help increase our detection rate were
not factored into the efficiency.

The above results apply when no significant amount of scattered light
was present.  Figures~\ref{fig:areamars} and \ref{fig:contour} show
the behavior of the scattered light near Mars.  Scattered light was
noticed up to about 20 arcminutes from Mars.
Figure~\ref{fig:effbright} shows the $50 \%$ R-band limiting magnitude
of the program versus the distance from Mars.

\section{Results and Discussion}

We found no new satellites of Mars to limiting apparent red magnitude
of 23.5 mag (Figure~\ref{fig:areamars}).  In order to determine the
corresponding size limit we relate the apparent red magnitude of an
object, $m_{R}$, to its radius, $r$, through
\begin{equation}
r = \left[ \frac{2.25\times 10^{16}R^{2}\Delta ^{2}}{p_{R}\phi
(\alpha)} \right]^{1/2} 10^{0.2(m_{\odot} - m_{R})}
\label{eq:appmag}
\end{equation}
in which $r$ is in km, $R$ is the heliocentric distance in AU,
$\Delta$ is the geocentric distance in AU, $m_{\odot}$ is the apparent
red magnitude of the sun ($-27.1$), $p_{R}$ is the geometric red
albedo, and $\phi (\alpha)$ is the phase function in which the phase
angle $\alpha=0$ deg at opposition.  For linear phase functions we use
the notation $\phi (\alpha) = 10^{-0.4 \beta \alpha}$, where $\beta$
is the ``linear'' phase coefficient.  If we use data from Table~1, a
phase coefficient of $\beta = 0.05$ mags per degree and assume the
albedo is 0.07 we find that an apparent magnitude of 23.5 corresponds
to objects that are about 0.09 km in radius at Mars' distance.  With
this kind of sensitivity we would have easily detected any Phobos ($r
\sim 11$ km) or Deimos ($r \sim 6$ km) sized satellites in our fields.
The survey covered $99.5 \%$ of Mars' Hill sphere.  Only the inner
$\sim 3$ arcminutes ($\le 0.07 r_{H}$) near Mars were not covered in
our survey due to scattered light from the planet.

Mars' lack of outer satellites compared to the giant planets is shown
in Figure~\ref{fig:irrsats} and Table~2.  By comparing Mars' Hill
sphere to the giant planets we would expect to find outer irregular
satellites much more distant than Phobos and Deimos.  Why are there no
known irregular satellites around the terrestrial planets?  Previous
authors have found that the inner terrestrial planets Mercury and
Venus probably have no satellites because of the consequence of strong
Solar tides which would make them unstable (Counselman 1973; Ward \&
Reid 1973; Burns 1973).  The Earth and Mars are spared extreme solar
tides.  The Earth might have lost small outer satellites due to the
presence of the large Moon (Callegari \& Yokoyama 2001).

There are two basic reasons why Mars may have no irregular satellites.
Either such objects are dynamically unstable on timescales comparable
to or less than the age of the Solar System or Mars lacked any
mechanism to capture irregular satellites in the first place.  The
proximity of Mars to Jupiter may limit distant stable orbits about
Mars.  The role of these perturbations on long timescales and as a
function of location within Mars' Hill sphere is still yet to be fully
explored.  Analysis of main-belt asteroid satellite stability shows
that the Sun is the dominant perturber for asteroids in the mid to
inner main belt while Jovian perturbations become dominant in the
outer main-belt (Chauvineau \& Mignard 1990).  Extrapolation to Mars
suggests that satellites may be stable for a significant portion of
the Hill sphere for the age of the Solar System.

Alternatively, the deficiency of irregular satellites around Mars may
arise from the difference in the formation process of Mars compared to
the giant planets.  Irregular satellite capture for the giant planets
probably occurred towards the end of the planet's formation epoch
where the energy dissipation needed for capture could be caused by gas
drag from an extended atmosphere, the enlargement of the Hill sphere
caused by the planet's mass growth and/or higher collision
probabilities with nearby small bodies (Kuiper 1956; Colombo \&
Franklin 1971; Heppenheimer \& Porco 1977; Pollack et al. 1979).
Unlike the giant planets, Mars is not expected to have gone through a
significant Hill sphere enlargement from a large mass accretion event
or to have had a dense extended atmosphere during formation.

\section{Conclusions}

No outer satellites of Mars larger than about 0.09 km in radius were
detected in our survey which covered $99.5 \%$ of Mars' Hill sphere
(from $3 \arcmin$ to over 0.7 degrees from the planet).  The absence
of outer satellites around Mars may indicate it did not have the right
conditions for irregular satellite capture as did the giant planets
during their formation or may show that such objects are dynamically
unstable, or both.

\section*{Acknowledgments}

We thank Jing Li for help on IDL contour plotting, Francesco Marzari
for a prompt review and the CFHT queue service observing team for
proficient data handling.  This work was supported by a grant to DJ
from NASA.  The Canada-France-Hawaii telescope is operated by the
National Research Council of Canada, Le Centre National de la
Recherche Scientifique de France, and the University of Hawaii.

\newpage

\begin{figure}
\caption{The area searched around Mars for satellites using the 3.6 m
CFHT telescope.  Four fields were imaged 3 times each on 2 separate
nights (UT July 5 and 7, 2003) for a total of 24 images.  The black dot
at the center represents Mars' position.  The dot is larger than the
apparent diameter of Mars and incorporates the orbits of Phobos and
Deimos.  The four different hatched squares represent the four fields
imaged around Mars on both nights.  Overlapping hatches show where
some of the fields overlapped in coverage.  The dotted circle shows
the Hill sphere of Mars to our line of sight.  The scattered light
from Mars was not a problem for most of the images but became quickly
dominant very near Mars.  The inner solid circles show where scattered
light became significant compared to the normal sky background.  At
the outer solid circle near 0.20 degrees the background was about
$1.5$ times higher, while at the middle solid circle the background
was about $2$ times higher and at the inner solid circle the
background was about $3.75$ times higher than the nominal sky
background.}
\label{fig:areamars} 
\end{figure}

\begin{figure}
\caption{Detection efficiency of the moving object search computer
program versus the apparent red magnitude.  The $50 \%$ differential
detection efficiency is at about 23.5 mags.  The differential
efficiency was determined by how many artificially generated moving
objects with motions similar to Mars were found by the program at a
given magnitude.  Magnitudes bins were separated by 0.1 mags.  This
efficiency is for no significant scattered light in the fields.  The
calculation of the effective radius assumes an albedo of 0.07.}
\label{fig:eff} 
\end{figure}

\begin{figure}
\caption{A contour plot of a survey image close to Mars to show the
scattered light.  The contours show $7000:6000:5000:4000:3000$ surface
brightness contours, where the nominal background sky level away from
the planet was near 2000 counts.  Mars is about 3 arcminutes south
from the bottom of the plotted region.  The vertical spike is caused
by diffraction from the telescope's support structure.  The image
color has been inverted so that higher counts are darker.  The image
color was also logged to bring out more detail between the top and
bottom of the image.}
\label{fig:contour} 
\end{figure}

\begin{figure}
\caption{$50 \%$ efficiency detection red limiting magnitude of the
survey versus distance from Mars.  Scattered light is only significant
starting at about 15 arcminutes from Mars. The calculation of the
effective radius assumes an albedo of 0.07.}
\label{fig:effbright} 
\end{figure}

\begin{figure}
\caption{A comparison of the known irregular satellites of the giant
planets with Mars' two known satellites.  The horizontal axis is the
ratio of the satellites semi-major axis to the respective planet's
Hill radius.  The vertical axis is the inclination of the satellite to
the equator of its planet.  Only Nereid of Neptune is close to the
position of the known Martian satellites in this parameter space.}
\label{fig:irrsats} 
\end{figure}

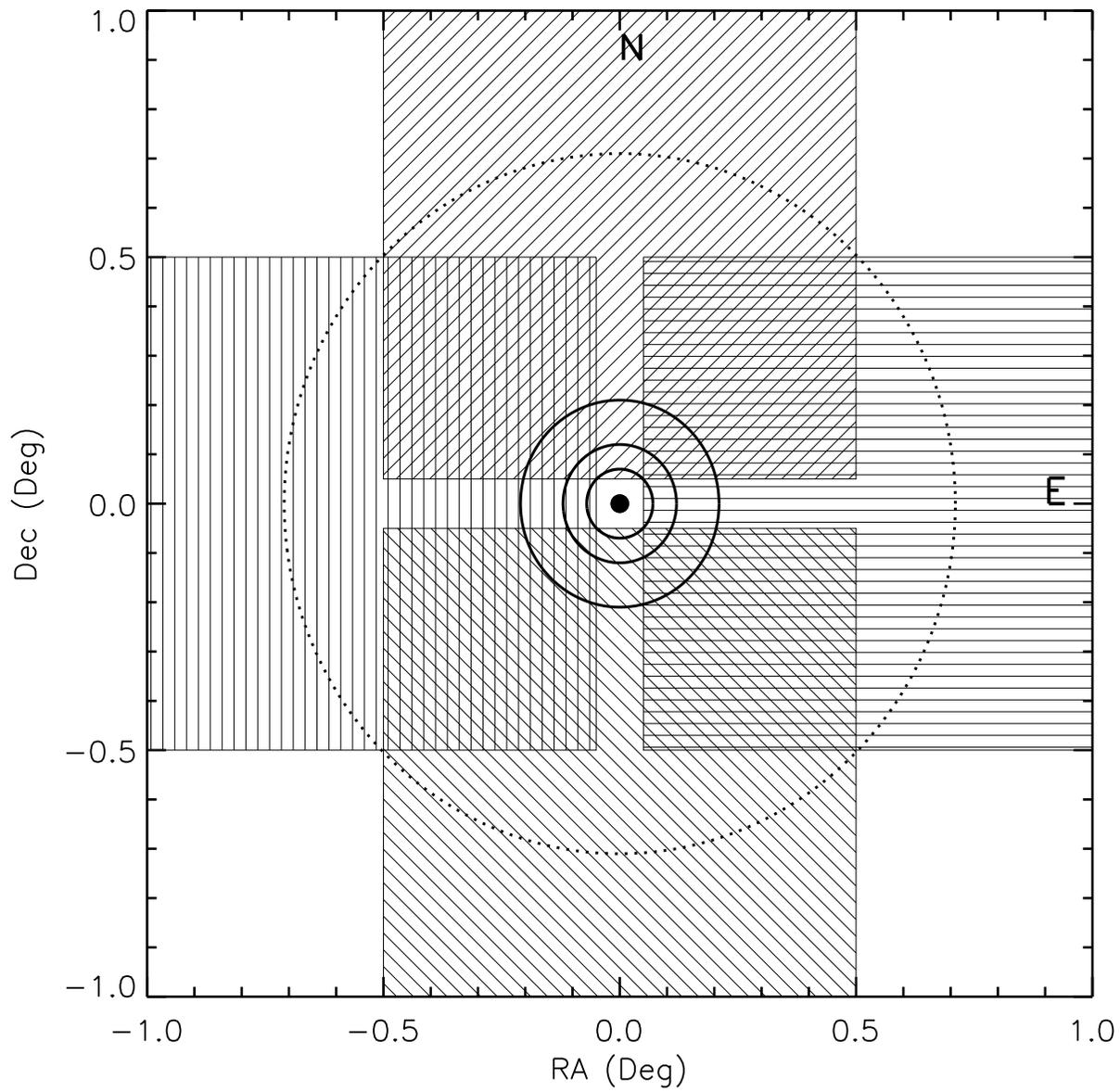
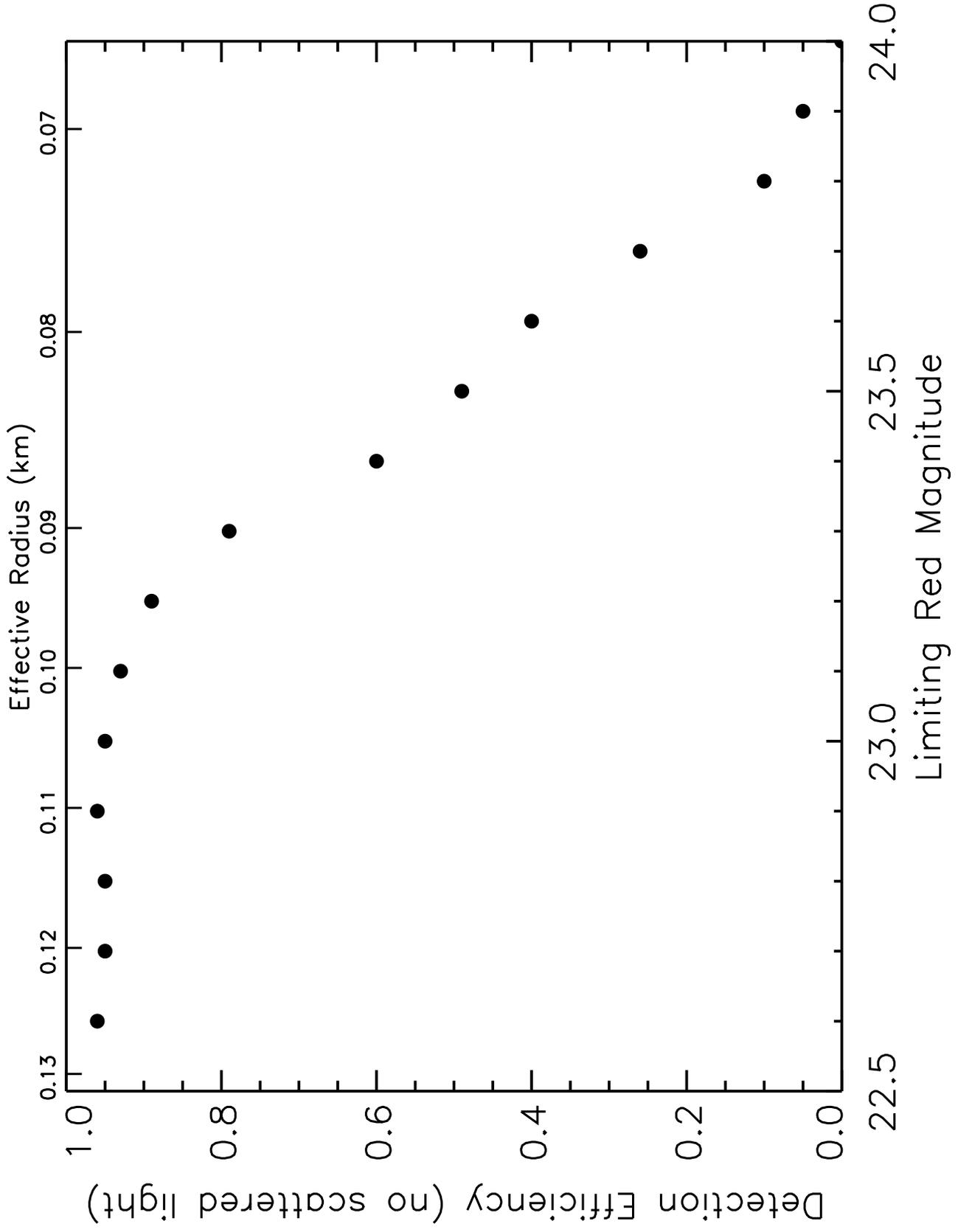
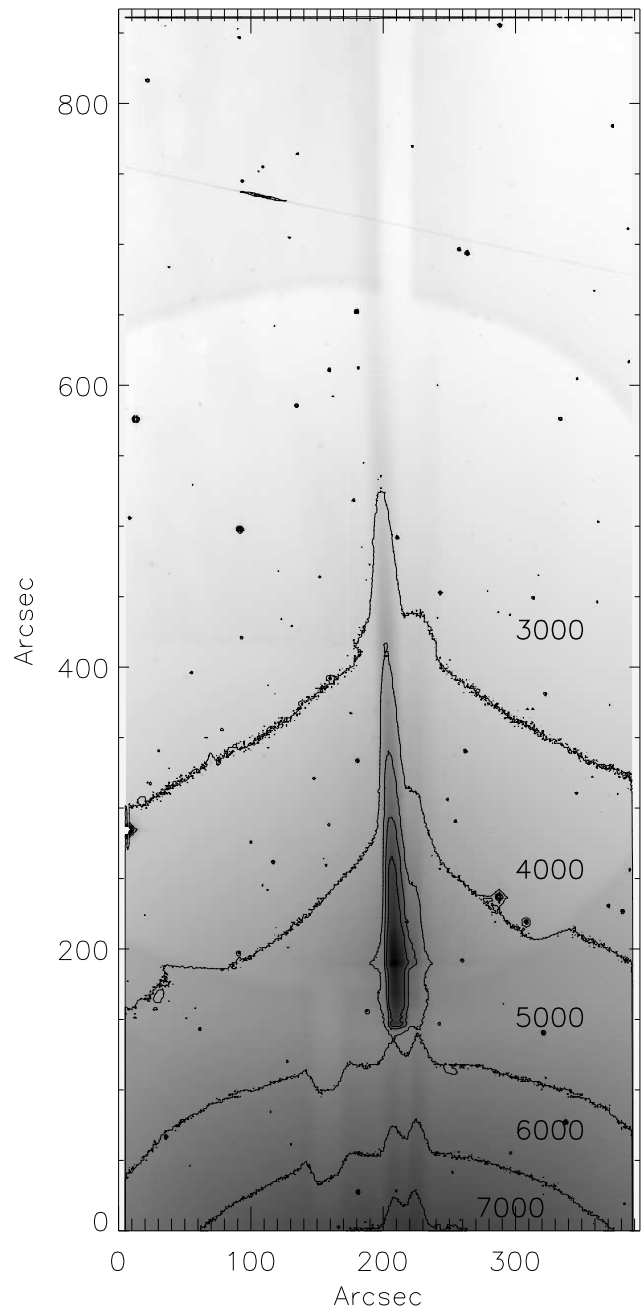
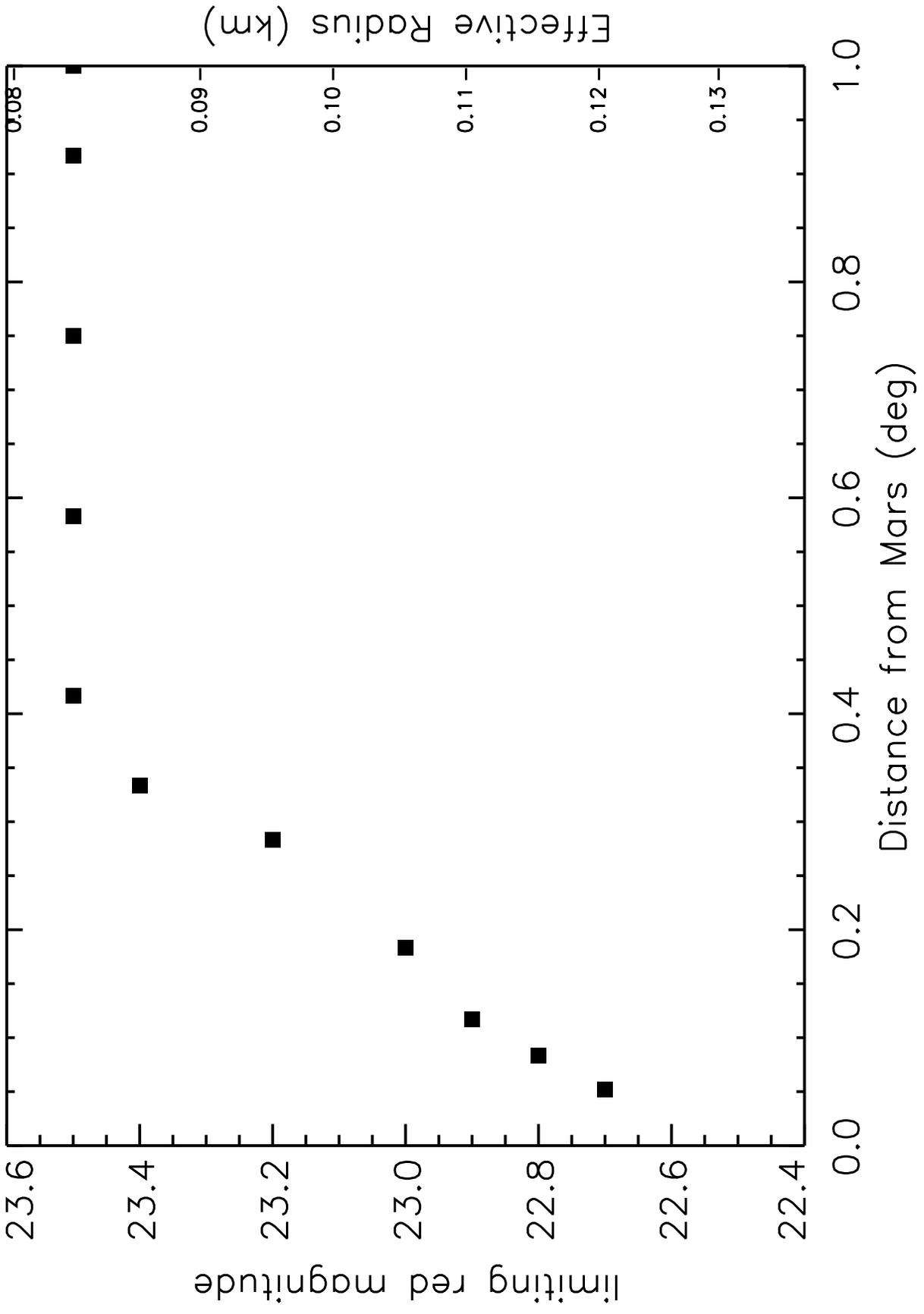
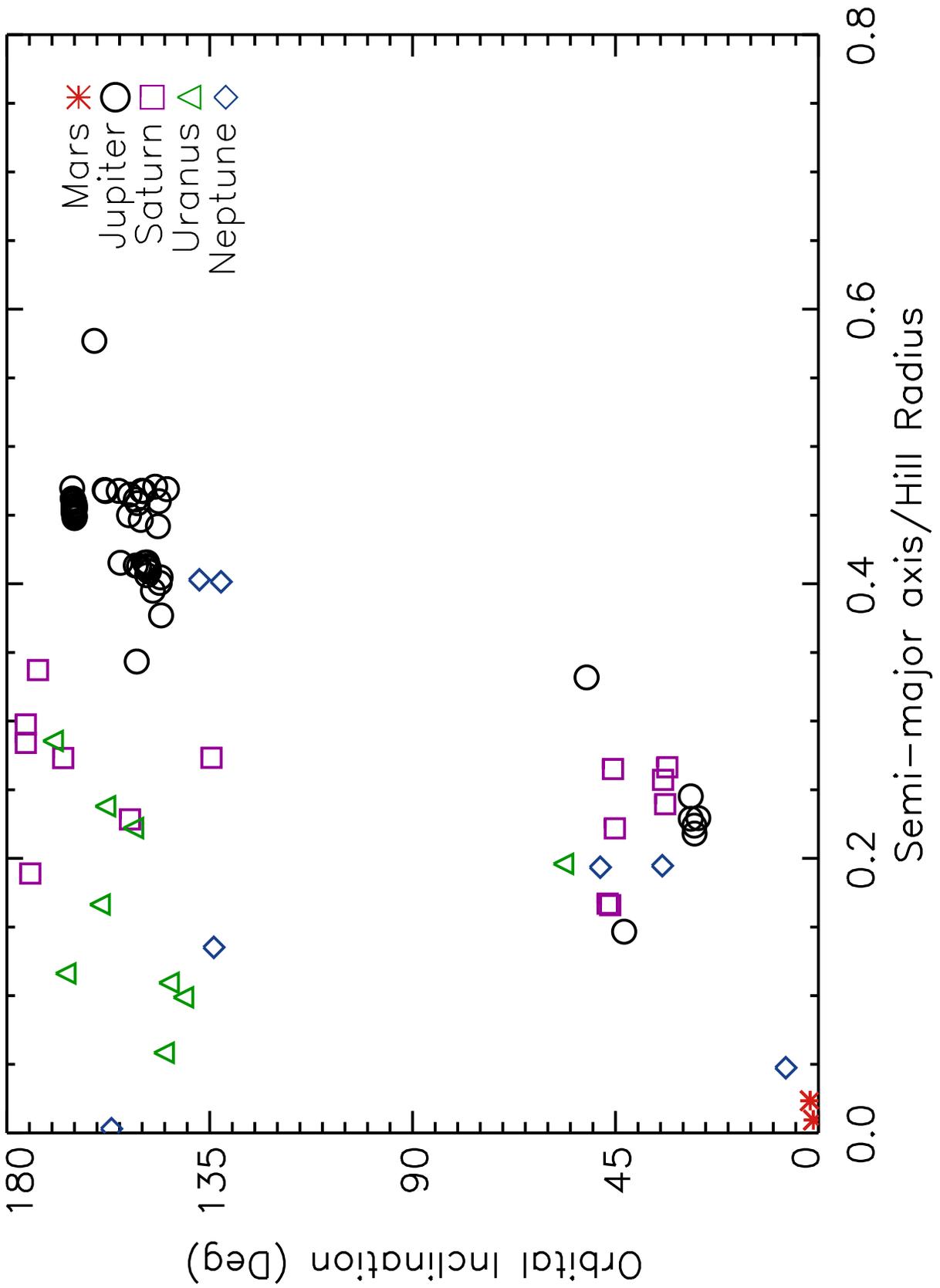
\end{document}